\documentstyle[11pt]{article}
\def\Journal#1#2#3#4{{#1} {\bf #2}, #3 (#4)}
\def\Preprint#1#2{{#1}{/#2}}
\def\NPB{{\em Nucl. Phys.} B}
\def\PLB{{\em Phys. Lett.}  B}
\def\PRL{\em Phys. Rev. Lett.}

\def\NPA{{\em Nucl. Phys.} A}
\def\PRC{{\em Phys. Rev.} C}

\begin{document}
\begin{titlepage}

\hspace{11cm}BIHEP-TH-97-50

\vspace{1cm}

\centerline{\large \bf H-DIHYPERON IN QUARK CLUSTER MODEL
\footnote{This work was partly supported by the National
Natural Science Foundation of China }}
\vspace{1cm}
\centerline{ P.N.Shen$^{a,b,c}$, Z.Y.Zhang$^{a}$, Y.W.Yu$^{a}$,
X.Q.Yuan$^{a}$, S.Yang$^{a}$}
\vspace{1cm}
{\small
{
\flushleft{\bf  $~~~$a. Institute of High Energy Physics, Chinese Academy
of Sciences, P.O.Box 918(4),}
\flushleft{\bf  $~~~~~~$Beijing 100039, China}
\vspace{8pt}
\flushleft{\bf  $~~~b.$ China Center of Advanced Science and Technology
 (World Laboratory),}
\flushleft{\bf  $~~~~~~$P.O.Box 8730, Beijing 100080, China}
\vspace{8pt}
\flushleft{\bf  $~~~c.$ Institute of Theoretical Physics, Chinese Academy
of Sciences, P.O.Box 2735,}
\flushleft{\bf  $~~~~~~$Beijing 100080, China}
\vspace{8pt}
}}

\vspace{2cm}

\centerline{\bf Abstract}

\noindent
The H dihyperon (DH) is studied in the framework of the SU(3) chiral
quark model. It is shown that except the $\sigma$
chiral field, the overall effect of the other SU(3) chiral fields is
destructive in forming a stable DH. The resultant mass of DH in a
three coupled channel calculation is ranged from 2225 $MeV$ to 2234 $MeV$.

\end{titlepage}

\baselineskip 18pt

In 1977, Jaffe predicted DH \cite{jaf}, a six-quark state with
strangeness $(s)$ being -2 and $J^{P}= 0^{+} (S=0, T=0$),
by using a simple color magnetic
interaction in the MIT bag model. Since then, many  theoretical
\cite{bal,yaz,sz,kod,wol,glo} and experimental \cite{ima} (and
references therein) efforts have been devoted to the DH study.
There were so many theoretical predictions of
DH which are quite different in different models. For instance, by using
the MIT bag model, Jaffe gave a binding energy of about $80MeV$ below
the two $\Lambda$ threshold ($E_{\Lambda\Lambda}$) \cite{jaf}; in terms
of the Skyrme model, Balachandran et al. even showed a larger binding
energy of about several hundred $MeV$ \cite{bal}; by employing the
cluster model, Yazaki et al. predicted the energy of DH from about 
$10MeV$ above $E_{\Lambda\Lambda}$ to about $10MeV$ below
$E_{\Lambda\Lambda}$
\cite{yaz,yaz1}; also, by using the cluster model but different
interaction with Yazaki's, Straub et al. announced a binding energy
around $20MeV$ \cite{sz}; in terms of Quantum Chromodynamics($QCD$)
sum rules, Kodama et al. gave a binding energy around $40MeV$ although
the error bar was quite large; considering mutually the two-cluster and
six-quark cluster configurations, Wolfe et al. obtained a deeply bound
DH with a binding energy about several hundred $MeV$ \cite{wol}; and
by employing a quark model without the one-gluon-exchange interaction (OGE),
Glozman et al. announced the non-existence of a bound DH \cite{glo}.
On the other hand, there is no experimental
evidence showing the existence of DH up to now. The provided lower limit
of the DH mass is about $2200MeV$ \cite{ima,ima1}.

The reason for carrying out such researches
is straightforward. According to the feature of the color
magnetic interaction (CMI) in the one-gluon-exchange potential (OGE),
when the strangeness of the system concerned is equal to -2, S=0 and
$(\lambda\mu)_{f}=(0~0)$, the expectation value of CMI
presents more attractive feature than those contributed by two
$\Lambda$ baryons. As a consequence,
six quarks could be squeezed in a small region, the typical short-range
$QCD$ behavior would be demonstrated and some new physics might be
revealed. Therefore, studying DH will be rather significant in
understanding the quark characteristics of the wave function of the
multiquark system and the short-range behavior of the $QCD$ theory.

On the other hand, till now, there still exist some uncertainties in the
nucleon-hyperon ($NY$) interaction on the baryon-meson degrees of freedom,
especially in the short-range part, so as the hyperon-hyperon ($YY$)
interaction and the prediction of DH. Therefore, the next generation, the
quark-gluon degrees of freedom, may be a more effective base to establish
the relations among the nucleon-nucleon ($NN$), $NY$ and $YY$ interactions,
and consequently to give a more reliable prediction of DH. As well known,
most nuclear phenomena are just of the low energy approximation of $QCD$.
There exist lots of nonperturbative $QCD~ (NPQCD)$ effects. Unfortunately,
nowadays one still cannot solve NPQCD directly,
and has to employs certain $QCD$
inspired models. The $SU(3)$ chiral quark model is just one of the most
successful ones. In that model, the couplings between chiral fields and
quark fields were introduced to describe the short- and medium-range
$NPQCD$ effects, and a more reasonable quark-quark interaction $V_{q-q}~~
(q=u,d,$ and $s)$\cite{zys} was obtained. With that $V_{q-q}$, one could
mutually describe the experimental $NN$ scattering phase shifts, available
$YN$ scattering cross sections \cite{zys} and some properties of
single baryons including the empirical masses of single baryon ground
states \cite{sdz,cz}. Extrapolating that model to the $s=-2$ system, one
can study the $YY$ interaction, double strangeness hypernuclei, DH and etc..
In this letter, we would choose DH as a target, because it is a simplest
case with two strangeness and its structure is relative simple so that
the interaction can be preliminarily examined by the present experimental
finding although it is only a lower limit of the DH mass, and the
short-range behavior can be revealed.

\vspace{0.3cm}

It is clear that both the DH structure and $V_{q-q}$ would affect the
theoretical prediction of the DH mass. Here, we first briefly introduce
the employed interaction. The Hamiltonian of a six-quark system in the
SU(3) chiral quark model reads

\begin{eqnarray}
H~=~T~+~\sum_{i<j}\big(V_{ij}^{CONF}~+~V_{ij}^{OGE}~+~V_{ij}^{PS}~+~
V_{ij}^{S}\big),
\end{eqnarray}
where $T$ denotes the kinetic energy operator of the system and
$V_{ij}^{CONF}$, $V_{ij}^{OGE}$, $V_{ij}^{PS}$ and $V_{ij}^{S}$
represent the confinement, one-gluon exchange, pseudo-scalar chiral field
induced and scalar chiral field induced potentials between the
i-th and j-th quarks, respectively. The confinement potential is
phenomenologically taken as

\begin{eqnarray}
V_{ij}^{CONF}~=~-(\lambda^{a}_{i} \lambda^{a}_{j})_{c}(a_{0~ij}
+a_{ij}r_{ij}^{2}),
\end{eqnarray}
which describes the long range nonperturbative $QCD$ effect. The introduced
$a_{0~ij}$ terms which take different values for different $q-q$ pairs are
called zero-point energy terms. They guarantee
that the empirical thresholds of
considered channels can more accurately be reproduced. The short range
perturbative OGE potential is chosen to be the commonly used form
\cite{yzs,yzy,zys,sdz}.
In order to restore the important symmetry of strong interaction, the
chiral symmetry, we introduce SU(3) chiral fields coupling to quark
fields so that the medium-range $NPQCD$ effects can be described
\cite{yzs,yzy,zys}. The pseudoscalar-field-induced potentials are:

\begin{eqnarray}
V_{ij}^{PS} = C(g_{ch},~m_{\pi_{a}},
~\Lambda)~\frac{m_{\pi_{a}}^{2}}{12m_{i}m_{j}}
\cdot f_{1}(m_{\pi_{a}},~\Lambda,~r_{ij})~(\vec{\sigma_{i}}\cdot
 \vec{\sigma_{j}})\cdot
(\lambda^{a}_{i}\lambda^{a}_{j})_{f}~,
\end{eqnarray}
and the scalar-field-induced potentials are:

\begin{eqnarray}
V_{ij}^{S} = - C(g_{ch},~m_{\sigma_{a}},
~\Lambda)  
 \cdot f_{2}(m_{\sigma_{a}},~\Lambda,~r_{ij}) \cdot
 (\lambda^{a}_{i}\lambda^{a}_{j})_{f}~,
\end{eqnarray}
where the subscript $f$ denotes that the operators in parentheses
are in flavor space. The expressions of $f_{i},~Y$, and $C$ are shown in
Ref.\cite{zys}. To retain the important chiral symmetry as much as
possible, we take all chiral-quark coupling constants to be the same value
\begin{eqnarray}
\frac{g_{ch}^{2}}{4\pi}~&=&~\frac{9}{25}~\frac{g_{NN\pi}^{2}}{4\pi}~
\Big(\frac{m_{q}}{M_{N}}\Big)^{2}.
\end{eqnarray}
In Eq.(3), $\pi_{a}$ with ($a=1,2,...,8,$ and $0$) correspond
to the pseudoscalar fields $\pi,~K,~\eta_{8}$ and $\eta_{0}$, respectively,
and $\eta$ and $\eta'$ are the linear combinations of $\eta_{0}$ and
$\eta_{8}$ with the mixing angle $\theta$. In Eq.(4),
$\sigma_{a}$ with ($a=1,2,...,8,$ and $0$) correspond to the  scalar
fields $\sigma',~\kappa,~\epsilon$ and $~\sigma$, respectively.

In solving this six-quark system problem, the first selected set of
model parameters is that used in Ref.\cite{zys}. This is because
that with this set of parameters, almost all empirical
partial wave phase shifts of the $N-N$ scattering can be well
re-produced, meanwhile the available cross sections of $N-Y$ processes
can reasonably be explained, some masses of baryon ground states can
accurately be obtained, and some properties of baryons such as $EM$
transition rates and etc. can better be understood \cite{cz,sdz}.
Thus, the predicted mass of DH is based  on a more solid ground,
and the reliability of the prediction is increased. Furthermore,
the possible range of the DH mass can be tested by shifting the
values of model parameters within reasonable regions.

\vspace{0.3cm}

Then, we show how to choose the model space
in solving the bound state problem of a six-quark system.
There are two types of possible configurations in studying
the structure of DH.

\vspace{0.3cm}

\noindent
{\em (1) Six-quark cluster configuration.}

In this configuration, the trial
wave function  can be expressed as the linear combination of differently
sized basis functions:
\begin{eqnarray}
\Psi_{(\lambda\mu)_{f}~T~S}~=~\sum_{i}~C_{i}~\Phi_{(\lambda\mu)_{f}~T~S}
(\omega_{i}),
\end{eqnarray}
with the basis function
\begin{eqnarray}
\Phi_{(\lambda\mu)_{f}~T~S} (\omega_{i})~=~\phi~[~(0s)^{6},~\omega_{i}]~
\chi^{f\sigma}_{(\lambda\mu)_{f}}~\chi^{c}_{(00)},
\end{eqnarray}
where $\phi~[~(0s)^{6},~\omega_{i}]$ is the orbital wave function
$(\omega=\frac{1}{mb^{2}})$, and $\chi^{f\sigma}_{(\lambda\mu)_{f}}$ and
$\chi^{c}_{(00)}$ denote the
wave functions in the flavor-spin and color spaces, respectively.
This trial wave function is in the pure symmetry-basis-function space. In
this configuration space, $(0s)^{6}$ configurations which describe the breath
mode are considered only.

\vspace{0.3cm}

\noindent
{\em (2) Two-cluster configuration.}

In this configuration, there exist three possible channels:
$\Lambda\Lambda$, $N\Xi$ and $\Sigma\Sigma$. In the framework of
Resonating Group Method (RGM), the trial wave function of DH can be
written as
\begin{eqnarray}
\Psi~=~\alpha\mid~\Lambda~\Lambda~\rangle~+~\beta\mid~N~\Xi~\rangle~+
~\gamma\mid~\Sigma~\Sigma~\rangle,
\end{eqnarray}
with the two-cluster wave function
\begin{eqnarray}
\mid~B_{1}B_{2}~\rangle~=~{\cal{A}}~\lbrack~\phi_{B_{1}}~\phi_{B_{2}}~
\chi_{rel}~{\cal{R}}_{cm}~\rbrack_{ST=00},
\end{eqnarray}
where ${\cal{A}}$ stands for the antisymmetrizer, $\phi_{B_{1~(2)}}$ denotes
the wave function of the cluster $B_{1~(2)}$, $\chi_{rel}$ represents the
trial wave function of the relative motion between clusters $B_{1}$ and
$B_{2}$ and ${\cal{R}}_{cm}$ is the wave function of the total center of mass
motion. This trial wave function is in the physics-basis-function space.
The physical picture of this configuration is that in the compound region
of two-interacting clusters (or composite particles), there might exist
a bound state or a resonance.

\vspace{0.3cm}

Let us define a quantity
\begin{eqnarray}
E_{H}~=~M_{H}~-~M_{\Lambda\Lambda},
\end{eqnarray}
where $M_{H}$ and $M_{\Lambda\Lambda}$ denote the mass of DH
and two $\Lambda$'s,
respectively. Apparently, $E_{H}~<~0$ stands for a stable DH
against weak decay.
$E_{H}$ (or $M_{H}$) can be obtained by solving the Schr\"{o}dinger equation
in which the above mentioned potentials are employed. The  results with
different configurations are discussed in the following.

\vspace{0.3cm}

\noindent
{\em (1) Six-quark cluster case.}

\vspace{0.3cm}

In this case, $\omega_{i}$ are taken as variational parameters. By using the
variational method, one can minimize the Hamiltonian matrix element with
respect to $\omega_{i}$. The resultant masses of DH are tabulated in Table I.

\newpage

\centerline{\bf {Table I. $~~~E_{H}(MeV)^{\dag}$ ~~in the six-quark
cluster case}}

\vspace{0.1cm}

%\begin{small}
\begin{center}
\begin{tabular}{|c|c|}
\hline
${~}$ & ${~}$  \\
${~~~~~}V^{OGE}~+~V^{CONF}{~~~~~~}$  &{~~~~~~~~~~} 311 {~~~~~~~~~~}\\
 &   \\
\hline
 &   \\
$V^{OGE}~+~V^{CONF}$  & 127 \\
+$V^{\pi}~+~V^{\sigma}$  &  \\
\hline
 &   \\
$V^{OGE}~+~V^{CONF}$  & 276 \\
+$V^{PS}~+~V^{S}$ &   \\
\hline
\end{tabular}

\vspace{0.1cm}

$\dag$ {\footnotesize {Parameters used are those in ref.\cite{zys}.}}\\
\end{center}

\vspace{0.1cm}

\noindent
It is shown that if one only employ OGE and confinement potentials,
the mass of six-quark cluster is $311MeV$ heavier than those of two
$\Lambda$'s, $M_{\Lambda\Lambda}$. When one additionally
employs $V^{\pi}$ and $V^{\sigma}$, the mass of DH
would decrease, but it still $127MeV$ heavier than $M_{\Lambda\Lambda}$.
However, when one includes all SU(3) chiral pseudoscalar and scalar
fields, namely employs the other chiral fields in additional to the
$\pi$ and $\sigma$ fields, the corresponding mass becomes larger again,
which is $276MeV$ heavier than $M_{\Lambda\Lambda}$. This means
that the couplings between $\sigma$ chiral field and
quark fields cause additional attraction, which is helpful to
reduce the mass of the six-quark system. On the contrary,
the overall effect of the
contributions from other SU(3) chiral fields provides a repulsive feature
so that  DH is quite hard to form. Moreover, no matter
in which case, the mass of DH is heavier than
$M_{\Lambda\Lambda}$, namely DH is not a stable particle against to
strong decay to $\Lambda\Lambda$ and/or $N\Xi$.
Therefore, a model space with $(0s)^{6}$ six-quark cluster structure only
may not be a favored model space in studying the DH structure.

\vspace{0.3cm}

\noindent
{\em (2) Two-cluster case.}

In RGM, to solve the bound state problem, one usually expands the unknown
relative wave function $\chi_{rel}$ by using locally peaked Gaussian basis
functions
\begin{eqnarray}
\chi_{rel}~=~\sum_{i}~c_{i}~\chi_{i}~,
\end{eqnarray}
where  $c_{i}~$'s are variational parameters \cite{kam}.

\vspace{0.3cm}

Due to the Pauli principle, there exists a forbidden degree  in the
six-quark trial function \cite{osy}
\begin{eqnarray*}
\Psi^{forbidden}~=~{\cal{A}}~\Big(~\frac{1}{\sqrt{3}}~\mid\Lambda\Lambda\rangle~+~
\frac{1}{\sqrt{3}}~\mid N\Xi\rangle~+~\mid\Sigma\Sigma\rangle~\Big).
\end{eqnarray*}
\noindent
This forbidden degree and almost forbidden degrees can be detected
by examining the zero and almost zero eigenvalues of
the normalization kernel, respectively \cite{wt}.
In particular, in the bound state RGM calculation,
a component with the inter-cluster distance to be zero in the
trial wave function, which is just a six-quark cluster configuration with
the [6] symmetry, has to be included so that the behaviors of
two clusters at the shorter inter-cluster distance can be
well described and the stable
and reliable solutions can be obtained. As a side-effect, the disturbances
from the forbidden and almost forbidden degrees become serious. Sometimes,
these disturbances would spoil the numerical calculation and produce
non-physical results. Therefore, only after all the non-physical
degrees are completely eliminated, the resultant energy of the bound
state can be trusted.
Due to the aforesaid reasons, at this moment, it may not be necessary to
further carry out the mixing of the configurations (1) and (2). Moreover,
in practice, eliminating the non-physical degree can be realized
by performing the off-shell transformation.
Then, carrying out the variational procedure,
one can obtain the mass of  DH or $E_{H}$.  The resultant $E_{H}$'s
are tabulated in  Table II.

\vspace{0.2cm}

%\newpage

\centerline{\bf {Table II. $~~~E_{H}(MeV)~~$ and ${\cal {R}}(fm)$
in the two-cluster case$^{\dag}$.}}

\vspace{0.2cm}

%\begin{small}
\begin{center}
\begin{tabular}{|c|c|c|c|c|}
\hline
%\cline{3-5}
\multicolumn{2}{|c|}{~} & ${~}$ & ${~}$ & ${~}$ \\
%\multicolumn{3}{|c|c|c|}{~} \\
\multicolumn{2}{|c|}{~} &
${\footnotesize {\big(~\mid\Lambda\:\Lambda\rangle~\big)}}$ &
 ${\footnotesize {\bigg(~{\begin{array}{cl}
      ~&\mid\Lambda\:\Lambda\rangle\\
     \!\!+\!\!&\mid N\Xi\rangle
        \end{array}}~
 \bigg)}}$ &
 ${\footnotesize {\Bigg(~{\begin{array}{cl}
     ~&\mid\Lambda\:\Lambda\rangle\\
     \!\!+\!\!&\mid N\Xi\rangle\\
     \!\!+\!\!&\mid\Sigma\Sigma\rangle
        \end{array}}~
 \Bigg)}}$ \\
\multicolumn{2}{|c|}{~} & ${~}$ & ${~}$ & ${~}$ \\
%\multicolumn{3}{|c|c|c|}{~} \\
\hline
 & & & & \\
 & $ E_{H}$ & 9.41 & 8.95 & 6.77\\
$V^{OGE}~+~V^{CONF}$ & & & & \\
\cline{2-5}
 & & & & \\
 & $\cal {R}$ & 1.89 & 1.85 & 1.68\\
 & & & &  \\
\hline
 & & & & \\
 & $ E_{H} $ & 5.47 & -38.71 & -65.80\\
$V^{OGE}~+~V^{CONF}$ & & & & \\
\cline{2-5}
+$V^{\pi}~+~V^{\sigma}$& & & & \\
 & $\cal {R} $ & 1.66 & 0.75 & 0.72\\
 & & & & \\
\hline
 & & & &  \\
 & $E_{H}$ & 4.39 & 4.02 & 1.98 \\
$V^{OGE}~+~V^{CONF}$ & & & & \\
\cline{2-5}
+$V^{PS}~+~V^{S}$ & & & & \\
 & $\cal {R}$ & 1.59 & 1.57 & 1.41\\
 & & & & \\
\hline
\end{tabular}

\vspace{0.2cm}

$\dag$ {\footnotesize {Parameters used are those in ref.\cite{zys} and
$\cal {R}$ denotes the root-mean-squared radius of DH.}}
\end{center}

\vspace{0.3cm}

\noindent
It is shown that in the first ($V^{OGE}+V^{CONF}$) and third ($V^{OGE}
+V^{CONF}+V^{PS}+V^{S}$) cases, the results do not support a bound DH.
In the second case ($V^{OGE}+V^{CONF}+V^{\pi}+V^{\sigma}$),
the one-channel calculation ($\mid\Lambda\Lambda\rangle$)
result ($E_{H}=5.47MeV$) also does not support a bound state, but, the two-
channel ($\mid\Lambda\Lambda\rangle+\mid N\Xi\rangle$) and three-channel
($\mid\Lambda\Lambda\rangle+\mid N\Xi\rangle+\mid\Sigma\Sigma\rangle$)
calculations show a bound state with the binding energies of 38.71MeV
and 65.80MeV, respectively. These results can roughly be explained by the
interaction matrix elements.
The contribution from $OGE$ shows repulsive feature in the $\Lambda\Lambda$
channel and attractive features in both $N\Xi$ and $\Sigma\Sigma$ channels,
and the net contribution from $\pi$ and $\sigma$ fields at the short distance
demonstrates the more attractive character in the $N\Xi$ and $\Sigma\Sigma$
channels than that in the $\Lambda\Lambda$ channel.
However, due to the existences of two strange quarks in the DH system,
the chiral clouds with strangeness surrounding interacting baryons
become important. Thus, in our opinion, all the SU(3) chiral fields
should be considered. In fact, after including these fields, the aforesaid
over-strong attraction disappears, and the results in all cases become
smooth. This phenomenon can be understood by the matrix elements of the
spin-flavor-color operators of chiral fields, namely $sfc$ coefficients,
which show that the contributions from $\sigma$ and $\eta_0$ present the
attractive character and those from the other mesons, i.e.,
$\pi$, $K$, $\eta_8$, $\sigma'$, $\kappa$ and $\varepsilon$, are repulsive.
Moreover, in the third case, although all coupled channel
calculations do not show a bound DH, in comparison with the result in the
six-quark cluster case, the inclusion of additional channel
would reduce the mass of DH. Eventually, the resultant mass of DH is
around the $\Lambda\Lambda$ threshold.

\vspace{0.3cm}

To further understand the calculated $E_{H}$, we also list the corresponding
root-mean-squared radii ($\cal {R}$) of $DH$ in
different cases in Table II. These
numbers indicate that in the two- and three-channel calculations, if one
considers $OGE$, $\pi$ and $\sigma$ only, the root-mean-squared radii
of $DH$ are $0.75 fm$ and $0.72 fm$ for the two and three coupled channel
cases, respectively. Therefore, $DH$ is a
bound state.  In all the other cases, the resultant
$\cal {R}$ values are greater than $1.4fm$, thus $DH$ is no longer bound.

\vspace{0.3cm}

How the values of major model parameters affect the resultant mass of DH
is also studied. Three coupled channel results are used as samples to
demonstrate these effects. When the width parameter $b$ increases,
the corresponding $E_{H}$ value changes to a smaller value. Decreasing the
mass of $\sigma$, would make the DH mass lighter.
Moreover, a smaller $s$ quark mass $m_s$ corresponds to a lower DH mass,
and decreasing the mixing angle $\theta$ would just lower the DH mass
in a very small amount. Among these parameters, the mass of $\sigma$ would
give the biggest effect on $E_{H}$.

\vspace{0.3cm}

Finally, we give the possible mass range of DH. As mentioned above, the $NN$
scattering phase shifts and the $NY$ scattering cross sections as well as
the mass of DH depend on the values of model parameters. Our calculation
showed that except the set of model parameters used in Ref.\cite{zys},
another set of model parameters, say $b=0.53fm,~~
m_{\sigma}=600MeV,~~m_{s}=430MeV,~~\theta=-23^{\circ},~~
\Lambda_{PS,~\sigma'}=987MeV,~~\Lambda_{\kappa,~\epsilon,~\sigma'}=1381MeV$,
which are almost the values in limits, can also fit the
experimental $NN$ and $NY$ data \cite{ysyz}. With this set of model
parameters, the resultant $E_{H}$ is $-6.9MeV$. Further extending these
parameter values to their physical limits with which the empirical $NN$ and
$NY$ data cannot even been reproduced, one can obtain the upper and
lower bounds of the DH mass. When $b=0.6fm,~m_{\sigma}=550MeV$,
the lower bound of $E_{H}$ is $-9.1MeV$. On the other direction, if one
takes $b=0.48fm,~m_{\sigma}=675MeV$, the upper bound of DH is
around $4.9MeV$.

\vspace{0.3cm}

From above calculations, one finds that in the framework of
our SU(3) chiral-quark model, as long as one picks up a set of model
parameters which satisfy the stability conditions, the masses of the
ground states of baryons and meanwhile can be used to fit the experimental
$NN$ and $NY$ scattering data, the resultant mass of DH would be rather
stable and would be ranged in a very small region. This mass is consistent
with the present experimental finding and  reflects that the SU(3)
chiral-quark model is reasonable.

\vspace{0.3cm}

As a summery, one may have following conclusions. The six-quark
system with strangeness being -2, $J^{P}=0^{+}$ (S=0, T=0) is
studied in two possible model spaces. One is in a six-quark cluster
configuration space with breath mode, and the other is in a two-cluster
configuration space with three possible channels. It is shown
that the $(0s)^{6}$ model space
is not larger enough even the breath mode is considered. Therefore, the
mass of DH in this model space is generally 100$\sim$300 MeV heavier than
that in the two-cluster model space. The similar result also appears in the
other six-quark system calculation, say Deltaron $(d^{*})$ \cite{yzy}.
In the two-cluster configuration case, the result shows
that the mass of DH is ranged from $2225MeV$ to $2234MeV$ if the
experimental $NN$ and $NY$ data should simultaneously be reproduced.
It seems that all the SU(3) chiral fields must be considered in
studying DH. In this case, the SU(3) chiral fields surround the
baryons and make the baryons more stable and
independent. Therefore, the interaction between two baryons becomes
weaker so that it is hard to form a stable six-quark particle.
It is also shown that the lower and upper bounds of the DH mass
in the SU(3) chiral quark model are
$2223MeV$ and $2237MeV$, respectively.

\vspace{0.3cm}
\noindent
{\bf Acknowledgement}

\vspace{0.3cm}

We would like to thank Prof. A.Faessler's fruitful discussion during the
Symposium Symmetry and Dynamics in Nuclear and Low Energy Particle Physics
in Blaubeuren-Tuebingen, Germany.

\end{document}